\documentclass[10pt,conference]{IEEEtran}
\IEEEoverridecommandlockouts
\usepackage[utf8]{inputenc}
\usepackage[T1]{fontenc}
\usepackage[english]{babel}
\usepackage{csquotes}
\usepackage[style=ieee,sortcites=true,sorting=none,maxbibnames=6, backend=biber,mincitenames=1,maxcitenames=2, dashed=false]{biblatex}
\usepackage{graphicx}
\usepackage{amsmath}
\usepackage{amssymb}
\usepackage{amsthm}
\usepackage{mathtools}
\usepackage{braket}
\usepackage{hyperref}
\usepackage{titlesec}
\usepackage{balance}
\usepackage{booktabs}
\usepackage{tabularx}
\usepackage{float}
\usepackage{tikz}
\usepackage{fontawesome}
\usepackage[skip=2pt]{caption}
\usepackage{subcaption}
\usepackage{fancyhdr}
\usepackage{float}
\usepackage{placeins}
\usepackage{cleveref}
\usepackage{algorithm}
\usepackage{algpseudocode}

\graphicspath{{plots}}
\addto\extrasenglish{

}

\newcommand{\code}[1]{$\textbf{\texttt{#1}}$}

\title{Creating Automated Quantum-Assisted Solutions for Optimization Problems \thanks{This project was supported by the Federal Ministry for Economic Affairs and Climate Action on the basis of a decision by the German Bundestag through the project \emph{Quantum-enabling Services and Tools for Industrial Applications (QuaST)}}}

\author{
    \IEEEauthorblockN{Benedikt Poggel\IEEEauthorrefmark{1}, Xiomara Runge\IEEEauthorrefmark{1}, Adelina Bärligea\IEEEauthorrefmark{1}, Jeanette Miriam Lorenz\IEEEauthorrefmark{1}}
    \IEEEauthorblockA{\IEEEauthorrefmark{1}Fraunhofer Institute for Cognitive Systems IKS}
     \{benedikt.poggel, jeanette.miriam.lorenz\}@iks.fraunhofer.de
}
\date{2025-05-26}

\begin{document}
\maketitle

\begin{abstract}
When trying to use quantum-enhanced methods for optimization problems, the sheer number of options inhibits its adoption by industrial end users. Expert knowledge is required for the formulation and encoding of the use case, the selection and adaptation of the algorithm, and the identification of a suitable quantum computing backend. Navigating the decision tree spanned by these options is a difficult task and supporting integrated tools are still missing. We propose the QuaST decision tree, a framework that allows to explore, automate and systematically evaluate solution paths. It helps end users to transfer research to their application area, and researchers to gather experience with real-world use cases. Our setup is modular, highly structured and flexible enough to include any kind of preparation, pre-processing and post-processing steps. We develop the guiding principles for the design of the ambitious framework and discuss its implementation. The QuaST decision tree includes multiple complete top-down paths from an application to its fully hybrid quantum solution.
\end{abstract}

\section{Introduction}

Many industries face complex optimization problems such as production planning, vehicle routing and many more. Classical algorithms typically either require prohibitively long computation times or deliver suboptimal solutions. Quantum computing (QC) has the prospect to reduce the number of computational steps needed or improve the solution quality in the future. However, the technology is not yet easily accessible to end users which hinders the adoption and further development of application-specific algorithms. Tools that lower the entry barrier and provide a level of abstraction are of great interest to a wide range of industrial applications. In particular, productive QC solutions need a close integration with classical compute power. However, venturing into quantum-assisted solutions requires a heavy investment for industries as specialists from various domains are required for the formulation, decomposition, encoding and algorithm tuning. Running quick tests with state-of-the-art methods rarely leads to satisfying results due to the complexity of the matter and the experimental stage of noisy quantum computing, which contrasts with the often miraculous claims by the hardware and software providers.

In the attempt to find useful QC applications, quantum optimization has developed into an active and dynamic research field. Before the advent of scalable error correction, the hope for better and faster solutions rests mainly on variational algorithms~\cite{cerezo_variational_2021} like the Quantum Approximate Optimization Algorithm (QAOA)~\cite{farhi_quantum_2014} or the Variational Quantum Eigensolver (VQE)~\cite{peruzzo_variational_2014}. Many variants, specializations and improvements have been proposed over the years~\cite{tilly_variational_2022, blekos_review_2024}. However, no practical quantum advantage for optimization problems~\cite{abbas_quantum_2023} could be proven and the application of quantum-assisted methods for industrial use cases is still in its infancy. In classical computing, mathematicians and computer scientists have worked decades to find extremely good approximate algorithms even for NP-hard problems, culminating in complex algorithms like the Concorde Solver for the Traveling Salesperson Problem (TSP)~\cite{applegate_david_l_traveling_2007}. Clearly, the bar for quantum-enhanced algorithms lies high. Every step of the solution needs to be highly connected, optimized, and adapted to the application. Experts from mathematics, physics, engineering and computer science are required to work together. Tools that simplify the adoption of QC need to be developed to foster the advancement of the utility of the technology for industrial applications.

Their design is no simple task: On the path from an application to quantum-assisted solutions, countless options need to be evaluated against each other. They span a large decision tree ~\cite{poggel_recommending_2023} whose complexity severely inhibits further research and the application of quantum-enhanced methods even in a prototypical stage.

The research project ``Quantum-enhanced Services \& Tools for Industrial Applications'' (QuaST) attempts to make QC-assisted solutions more accessible to end users and allow researchers to test their methods on relevant application cases. As a central result of QuaST, we propose an integrated software tool that bundles existing knowledge in a modular and flexible way, and ultimately allows to construct an application-facing abstraction layer that can act as the top layer of any QC software stack. The concept and groundwork has been laid out in~\cite{poggel_recommending_2023}. The underlying structure of the automation engine is a directed acyclic graph of options that are weighed against each other, called the QuaST decision tree (QDT). Its goal is to bring QC to the end users. In return, it also allows researchers to connect with practitioners from logistics, finance and the many other industry areas where optimization problems are relevant.

Our main contributions are 
\begin{itemize}
     \item the identification of central design principles for a framework working towards accessible quantum-assisted optimization, 
     \item a discussion of our prototypical implementation of that framework, 
     \item the concept for the expansion of its features aiming for a useful application-facing abstraction layer. 
\end{itemize}
The QDT is a starting point for applied QC, simplifying the access for end users without expert knowledge in QC. It is highly modular and flexible to allow researchers to integrate new state-of-the-art methods and link them to use cases. 

The prototypical implementation we provide is written in Python and relies on Qiskit~\cite{javadi-abhari_quantum_2024}, but can be extended with interfaces to any other language. It guides the user through the process of solving an optimization problem with quantum-enhanced methods like VQE or QAOA, starting with the generation of the problem to the selection of a simulated or real quantum backend. The capacities of the QDT are growing, both in the number of accessible options (the ``size'' of the QDT) and its features. A shortlist of planned features is given in~\cref{sec:outlook}. End users can start their quantum journey at the point they are familiar with their application domain. Researchers can evaluate their methods in various setups, bringing quantum technology and its users closer together.

The remainder of this paper is organized as follows: Related efforts in automating the solution of optimization problems with quantum-enhanced methods are presented in \cref{sec:related}. The QDT guiding principles are derived in modifying the previous work from~\cite{poggel_recommending_2023} are laid out in \cref{sec:systematics}. \Cref{sec:implementation} then describes the functioning and structure of the QDT implementation including its building blocks and how they interact with each other. The concept on how to use the decision tree as an end user, or build on it as a developer or researcher is described in \cref{sec:demo}. Finally, our contributions are summarized, discussed and an outlook is given in \cref{sec:outlook}.

\section{Related Work}
\label{sec:related}

Despite the growth in research around quantum-assisted methods for optimization problems (at the time of writing, an arXiv search for the terms ``quantum computing optimization'' in paper abstracts yields roughly 1500 papers over the last 12 months), many researchers focus on very specific questions such as algorithmic advancements or using a specific method on a standard optimization problem. With the current state of QC and the open question of quantum advantage, progress towards abstraction layers gets a lot less attention (a similar arXiv search finds around 50 papers, with many false positives). In particular, abstraction layers typically concentrate on the compilation and transpilation steps of quantum circuits and omit the problem of designing the right algorithm and quantum circuit starting from an application. Nonetheless, some researchers have shed some light on the topic from different perspectives.

Matteo et al.~\cite{matteo_abstraction_2024} propose a concrete architecture from a computer scientist's view. Towards the application, the idea of a decision tree that aids in building quantum-assisted solutions is layed out in~\cite{poggel_recommending_2023}. Both publications point out that taking good decisions (e.g., circuit optimizations or problem formulations) is difficult. A central obstacle is the strong dependence between the decisions, which makes it hard to break the process down into manageable parts.

The present QuaST decision tree is an enhancement of the layered structure presented in~\cite{poggel_recommending_2023}. Although we drop the strict notion of layers in favor of the more versatile \emph{blocks} (see \cref{sec:systematics}), our implementation is inspired by the layers identified there: problem formulation, encoding, algorithm selection, selection of the classical optimizer (if necessary) and finally compiler and backend selection. The latter lie in the focus of various automation efforts~\cite{matteo_abstraction_2024, quetschlich_mqt_2023, henderson_demonstration_2024, bandic_full-stack_2022}. For this reason, the decision tree only treats it on a fundamental level, leaving the details to the specialized tools. 

Concepts for the automated selection of quantum-enhanced solvers exist building on machine learning~\cite{volpe_predictive_2024} or as an integrated form requiring user selections~\cite{volpe_towards_2024}. The systematic exploration of solution paths known as meta-solving has also been applied to quantum computing in a proof-of-concept way~\cite{eichhorn_hybrid_2024}.

Of course, the research community around high-performance computing knows the task of handling user input by end users that are relatively unfamiliar with the technology. With the integration of quantum hardware on the horizon, they have started demonstrating the full-stack workflow connecting quantum and HPC systems~\cite{bieberich_bridging_2023} and have proposed a software architecture~\cite{saurabh_conceptual_2023}. Outside an HPC context, an open-source cloud-based resource manager has been designed~\cite{chakraborty_empowering_2024} that is supposed to enlarge the potential user base of quantum computing.

Further automation efforts are concerned with the generation of quantum circuits, both for fault-tolerant quantum computing ~\cite{mao_q-gen_2024} and QAOA~\cite{farhi_quantum_2014} mixers~\cite{kulshrestha_qarchsearch_2023}.

Finally,~\cite{rohe_problem_2024} proposes a solution pipeline for optimization problems with quantum hardware, but focuses on the development process. Five stages are identified (use case identification, solution draft, pre-processing, execution, post-processing) as a guideline for developers on how to tackle optimization use cases. These stages are not related directly to the layers in~\cite{poggel_recommending_2023}, but can be read as a project template for manually developing quantum solutions for optimization problems.
\section{Systematic Approach to Quantum-Assisted Solutions for Optimization Problems}
\label{sec:systematics}

To begin with, automation requires a systematic understanding. For the purpose of the QuaST decision tree, a first step has been made in~\cite{poggel_recommending_2023}. There, five layers have been identified: problem formulation, encoding, algorithm selection, classical optimizer selection, and compilation/hardware selection. Hardware control, compilation and transpilation have been simplified to one layer with the plan to use existing tools and efforts for good compilation passes etc. The focus of~\cite{poggel_recommending_2023} as of this paper lies on the application side of the quantum-assisted solution.

\begin{figure}[ht]
    \centerline{\includegraphics[width=1.05\columnwidth]{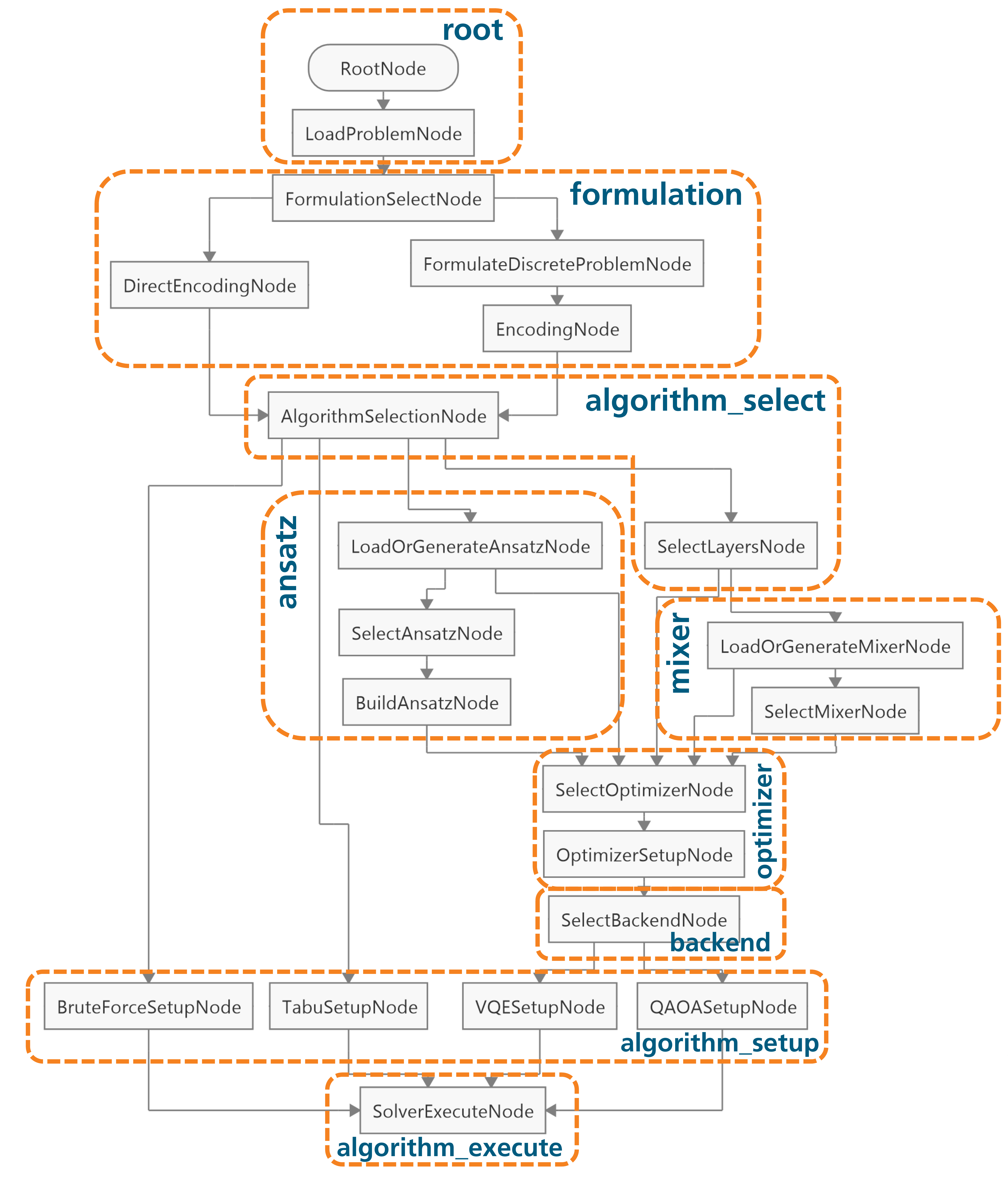}}
    \caption{The implemented nodes of the current version of the decision tree. The structure roughly follows the layered approach from~\cite{poggel_recommending_2023} where the exact setup varies for each algorithm.}
    \label{fig:decisiontree}
\end{figure}

A strictly layered approach certainly has its limits as pointed out in~\cite{poggel_recommending_2023}. From the point of view of the QuaST decision tree, the main shortcoming is the focus on variational quantum algorithms (VQA) like the Quantum Approximate Optimization Algorithm (QAOA)~\cite{farhi_quantum_2014} or the Variational Quantum Eigensolver (VQE)~\cite{peruzzo_variational_2014}. These algorithms have a specific structure in how they need to be set up (in particular regarding the involvement of a classical optimization algorithm). The QuaST decision tree in its current form also includes other algorithm types, for instance for classical comparison (brute-force sampling,  Tabu sampling~\cite{palubeckis_multistart_2004}). Even state-of-the-art VQAs often have a more complex relationship between classical and quantum resources (e.g. intermediate ``filter'' processing steps~\cite{amaro_filtering_2022} or adaptive variants~\cite{turati_benchmarking_2023}). To allow this flexibility, the QuaST decision tree does not arrange its nodes in layers, but rather connected \emph{blocks} of connected nodes. In \cref{fig:decisiontree}, these are denoted as orange boxes. Blocks share the characteristic of layers that once exited, a block cannot be reentered (unless by doing a partial restart to correct some error). However, they can be arranged more freely and, with their connections, form a directed acyclic graph. On the implementation level, these blocks have a one-to-one connection to the files of the centerpiece of the QuaST decision tree, the \code{decisiontree/interactive} subpackage. Its implementation is described in detail in \cref{sec:implementation}.

The QuaST decision tree has to ensure that it is accessible and useful to end users and researchers. This is reflected in a set of guiding principles:

\begin{enumerate}
    \item \emph{Modularity}: The insertion and removal of new submodules should be easy.
    \item \emph{Automation without Restriction}: The decision tree should allow expert users to make their own decisions if they wish so, but provide automated recommendations to other users.
    \item \emph{Locality}: Whenever possible, nodes should only communicate with their direct neighbors. Only direct neighbors should need to know implementation details of each other (if at all).
    \item \emph{Configurability}: Different users should be able to configure both whether a node is executed and how it is executed through a central configuration file.
    \item \emph{Light-Weight Execution}: Decision tree nodes should attempt to reduce computational overheads.
    \item \emph{Transparency}: Nodes should be transparent in how they modify the problem data.
\end{enumerate}

These guiding principles are no axioms and subject to continuous review and adaptation. Nonetheless, they form the vision of the QuaST decision tree.

Besides the actual tree structure described in \cref{sec:implementation}, the components of the decision tree can be used separately as well. Auxiliary functions and classes have been arranged into five subpackages.

\begin{itemize}
    \item \code{algorithms}: Contains both classical and hybrid quantum algorithms implemented with the \code{Solver} interface (see \cref{sec:implementation} for details.
    \item \code{evaluation}: Contains tools for evaluating and monitoring the convergence and behavior of variational quantum algorithms (loss landscape plotting and training path plotting). Currently not accessible through the interactive version.
    \item \code{mapping}: Contains conversion and utility functions to handle various bits of information related to optimization problems. In a future version, this might be merged with \code{utils}.
    \item \code{problems}: Contains the problem classes for included optimization problem and their abstract parent class \code{OptimizationProblem}
    \item \code{utils}: Contain general utilities, currently only the builder and hyperparameter classes described in \cref{sec:builder}
\end{itemize}

The primary use of these subpackages is to provide the code base for the decision tree. However, their implementation naturally advances the standardization of QC solutions by providing missing pieces necessary for its automation.
\section{Implementation}
\label{sec:implementation}

This section describes the concrete implementation of the QDT and serves as an entry point and reference guide for users and developers alike. It is mainly concerned with the framework that allows the QDT to function, with its building blocks and their relation to each other. For some examples on how to use the decision tree or contribute to it, see \cref{sec:demo}. In the remainder of this section, \Cref{sec:workflow} describes the algorithm behind the decision tree. \Cref{sec:building_blocks} gives more detail on its building blocks: nodes, queries, builders and hyperparameters. Then, \cref{ref:specific_layers} shows how specific nodes can be automated provided they find the correct objects to work with (e.g., optimizers).

For this section, it is expected that the reader has a basic knowledge of object-oriented programming and the Python data types. In particular, Python dictionaries are used as a versatile and flexible way to manipulate and hand over various types of data.

\subsection{Basic Algorithm}
\label{sec:workflow}

The basic algorithm of the QDT is simple to understand: It executes code from a series of \emph{nodes} that are arranged as a directed acyclic graph. The nodes modify a common dictionary called \code{problem\_data}. In doing so, they perform all necessary tasks and are, in principle, unlimited in what they can do. There are QDT nodes to load the problem, encode it, request input from the user and so on. For the latter, the QDT provides various \code{query} objects described in \cref{sec:queries}. The QDT package can be simply executed via \code{python -m decisiontree} which triggers the execution of a first node (printing a welcome message) and continues with problem loading. The behavior of the nodes can be altered with a \code{config.json} file at the project root. It is used to set levels of automation, specify locations to store data, and save access tokens for proprietary hardware and software.

On a path from the top of \cref{fig:decisiontree} to the bottom (the \emph{forward pass} or \emph{solution path}), the QDT executes nodes until it arrives at a final node. The final node triggers the execution of the algorithm and waits for the result. Subsequently, the \emph{backward pass} is performed. The result is passed through the solution path in reverse order. Each node can then interpret and modify the result based on its modification of the problem data beforehand. For example, the objective value can be corrected by a constant offset. More generally, decoding and any kind of post-processing takes place in the backward pass. When it reaches the top again (specifically \code{LoadProblemNode}, it saves result, problem instance, problem data and configuration file. With these files, the QDT solution paths become reproducible and allow for any kind of debugging or result analysis.

The pseudocode of the QDT is given in \cref{algo:decisiontree}. It describes a mere framework with any actual transformations of the problem data being left to the nodes. However, it provides a fully flexible structure to manipulate the problem data of an optimization problem in any way, to build quantum-enhanced solutions, and to interpret their results. For a more concrete example of a node, see \cref{algo:loadproblemnode} for the loading or generation of an optimization problem. While user input can be collected, it is possible to configure the QDT such that it creates and implements recommendations for, e.g., the best algorithm to run depending on problem class or instance characteristics. In this operational mode, the complexity of the decision tree is shielded from the end user, resulting in an abstraction layer from the problem instance to the algorithm.

\begin{algorithm}
\caption{QuaST Decision Tree}\label{algo:decisiontree}
\begin{algorithmic}[1]
\State config $\leftarrow$ load\_config()
\State node $\leftarrow$ init\_root\_node()
\State path $\leftarrow []$
\State problem\_data $\leftarrow$ init()
\While {node is not final}
    \State problem\_data $\leftarrow$ node.execute(problem\_data)
    \State append node to path
    \State node $\leftarrow$ node.next\_node()
\EndWhile
\State result $\leftarrow$ node.execute(problem\_data)
\For {node in reverse(path)}
    \State result $\leftarrow$ node.interpret\_result(result)
\EndFor
\end{algorithmic}
\end{algorithm}

\subsection{Building Blocks}
\label{sec:building_blocks}

The central building blocks of the decision tree are nodes, queries and builders. Nodes implement the core functionality and are the basic elements of the decision tree. Queries help interact with the user and builders allow automating the creation of solvers, quantum circuits etc. (in general: Python classes). This automation is a central benefit of the QDT since it bundles an underlying object together with the hyperparameters that define it.

\subsubsection{Queries}
\label{sec:queries}

Nodes frequently use queries to gather user input. Put simply, a \code{Query} is a class with an \code{input} function. It typically presents some options to the user or specifies a required data type. Once the user has provided input, it checks its validity and repeats the query if necessary. The \code{Query} class itself is abstract, but there are concrete subclasses designed to handle various types of inputs as listed in \cref{table:queries}. A special feature of the \code{Query} base class is the ability to deal with default values. Roughly spoken, whenever a user inputs an empty string, a query replaces it with a default value (if one exists). Furthermore, queries by convention can be answered by typing ``exit'' to abort execution of the decision tree immediately. 

\begin{table*}
\centering
    \begin{tabularx}{\textwidth}{llc}
         \toprule
         Query Class & Description & Parent \\
         \midrule
         \code{Query} & Abstract Base Class with concrete logic for handling default values \\
         \code{MultiChoiceQuery} & Allows user to select one of multiple options & Query \\
         \code{StringQuery} & String input & Query \\
         \code{PathQuery} & A StringQuery with optional checking whether the input is a path pointing to an existing file or directory & StringQuery \\
         \code{IntQuery} & Integer input & StringQuery \\
         \code{FloatQuery} & Float input & StringQuery \\
         \code{HyperParamQuery} & Requests a value for a hyperparameter (see \cref{sec:builder}) & Query \\
         \code{AbortQuery} & Asks whether the execution of the decision tree should be aborted & Query \\
         \bottomrule
    \end{tabularx}
    \caption{Query types included in the decision tree.}
    \label{table:queries}
\end{table*}

Finally, queries can be combined into \emph{query trees}. They do not only collect queries in one object, but allow for their conditional execution. For instance, some queries of the \code{LoadProblemNode} (\cref{algo:loadproblemnode}) depend on how the first query was answered.

\subsubsection{Nodes}
\label{sec:nodes}

Nodes are the core elements of the decision tree. They are executed in the loop shown in \cref{algo:decisiontree}. A valid implementation needs to provide two functions: the \code{execute} and \code{next\_node} routines. 

\code{execute} performs the task given to the node by manipulating the \code{problem\_data} dictionary. Furthermore, it has access to the configuration values. Its return value, a dictionary \code{path\_info}, encapsulates local information. It is only used to determine the next node and influence the backward pass execution of its own node. The \emph{next\_node} function returns the node the QDT should execute next, possibly deciding on the basis of \code{path\_info}.

On the backward pass of the decisiontree, the \code{interpret\_result} function of the node is executed. Implementing this function is optional. The default behavior inherited from the abstract base class does not modify the result dictionary. When called, it is passed all information from the problem data, configuration, \code{path\_info} and the result. Some nodes interpret the result by extracting specific information (such as a computational basis state) from it, others may perform postprocessing tasks like shifting the objective function towards a meaningful application measure. 

Nodes are universal in the tasks they can perform and the ways they modify the problem data. For a concrete example, see \Cref{algo:loadproblemnode}. The \code{execute}, \code{next\_node} and \code{interpret\_result} functions for the node loading an optimization problem instance. The execute function asks the user whether a problem instance should be loaded or generated randomly, and subsequently loads a problem instance or generates one at random. \code{next\_node} returns a node allowing to select the formulation for a given optimization problem (with the available options depending on the problem class) whereas \code{interpret\_result} saves the problem data (effectively the last step of the backward pass).

While \code{LoadProblemNode} is a simple node, it highlights basics like querying the user, referring the QDT to the next node, and performing a decoding task on the backward pass. In fact, many nodes are not more complicated. A major advantage of the QDT is its modularity which allows the single modes to perform only small, simple tasks.

\begin{algorithm}
\caption{LoadProblemNode}\label{algo:loadproblemnode}
\begin{algorithmic}[1]
\renewcommand{\algorithmicrequire}{\textbf{Input:}}
\renewcommand{\algorithmicensure}{\textbf{Output:}}
\Function{execute}{problem\_data, config}
    \State query: load or generate?
    \If {load}
        \State query: file path?
        \State instance $\leftarrow$ load\_from\_file()
    \ElsIf {generate}
        \State query: problem class?
        \State query: problem size?
        \State instance $\leftarrow$ generate\_random\_instance()
        \State save\_instance\_to\_file(instance)
    \EndIf
    \State problem\_data $\leftarrow$ insert(instance)
    \State path\_info $\leftarrow$ [instance] \Comment{available formulations depend on the problem instance}
    \State path\_info
\EndFunction
\Function{next\_node}{path\_info}
    \State return FormulationSelectNode(path\_info[instance])
\EndFunction
\Function{interpret\_result}{result, problem\_data, config, path\_info}
    \State save\_all\_data()
\EndFunction
\end{algorithmic}
\end{algorithm}

\subsubsection{Builders and Hyperparameters}
\label{sec:builder}

The research and development of the QuaST decision tree revealed a central obstacle for the automated creation and selection of various elements like optimizers or parameterized quantum circuits. The hardware-facing interfaces are standardized (e.g., a classical optimizer has to provide a \code{minimize} function within Qiskit). However, the user-facing interface is not. Initializing an optimizer in Qiskit requires reading the documentation to find the hyperparameters defining its behavior. Templates that create parameterized quantum circuits (PQC) to use as ansatz are in a similar state. For this reason, the QDT employs versatile objects called \emph{builders}. A builder is a wrapper for an underlying Python class that keeps track of the hyperparameters needed to create an instance of the underlying class. Instead of working with the underlying class directly, the QDT nodes can work with the corresponding builder. Most importantly, they can retrieve information about its hyperparameters in a standardized fashion. In a manual run, the QDT will list the hyperparameters and request values for them from the user. Most importantly, it can recommend values, either by setting a default or by invoking more sophisticated hyperparameter optimization methods. Finally, the builder can fix the hyperparameter values and create the corresponding instance of the underlying class.

The builder base class works with any Python class. Specialized builder classes are included in the QDT that keep track of their instances. Thus, nodes can create instances of specific object types (e.g., optimizers) easily. In the current decision tree version, specialized builders are used for the classical optimizer, variational ansatzes (VQE) and mixer (QAOA).

A builder thus contains an underlying class together with its hyperparameters. Hyperparameters are defined by the properties in \cref{table:hyperparams}. Besides the name and type, properties customize the behavior of the hyperparameter, most importantly what values it accepts.

\begin{table*}
\centering
    \begin{tabularx}{\textwidth}{lp{14cm}c}
         \toprule
         Property & Description & Default \\
         \midrule
         \code{name} & The name of the hyperparameter &  \\
         \code{description} & An optional description presented to the user if querying a value. & None \\
         \code{type} & The type of the hyperparameter (e.g., integer). &  \\
         \code{default} & An optional default value for the hyperparameter & None \\
         \code{test} & Optional additional tests to determine the validity of a proposed value for the hyperparameter (e.g., it must be larger than zero) & None \\
         \code{allow\_multiple} & Whether the hyperparameter also accepts lists of values instead of a single value. E.g., circuit layers that may consist of one or multiple gates. & False \\
         \bottomrule
    \end{tabularx}
    \caption{The properties of a hyperparameter.}
    \label{table:hyperparams}
\end{table*}

\subsection{Specific Nodes}
\label{ref:specific_layers}

Some QDT nodes try to identify all objects of a certain type by searching for subclasses of an abstract base class such as \code{OptimizationProblem}. As is common in object-oriented programming, the subclasses need to provide a specific interface. For problem classes and solvers, it is described in the following subsections.

\subsubsection{Problem Classes}
\label{sec:problemclass}

A problem class compatible with the QDT needs to implement the following methods:

\begin{itemize}
    \item \code{create\_random\_instance}: The ability to create random instances, given just the problem size as input, is currently required for quick testing of the QDT.
    \item \code{from\_dict}: The function reading data from a dictionary and creating a problem instance from it is invoked whenever a problem is loaded from a JSON file. New problem classes define their own standard for necessary and optional dictionary items. Problem instances should be able to be defined with the minimum information required. E.g., a TSP does not need city coordinates, but only the distances between the cities.
    \item \code{evaluate\_objective}: Given an algorithm result, the higher-level nodes of the QDT need to compute the corresponding objective value. Importantly, the application-specific performance measure (e.g., a path length) should be computed by this function.
    \item \code{to\_discrete\_problem} or \code{formulate\_problem}: At least one of these functions needs to be implemented. The discrete problem approach is slower, but can give immediate access to many encodings. For efficiency, \code{formulate\_problem} is preferred since it can be optimized in problem-specific ways. The latter also needs the developer to set the property \code{direct\_encoding\_nodes} with a list of supported encoding modes (e.g., ``one-hot'' and ``binary'').
    \item \code{to\_dict} is not required, but strongly advised as the inverse function to \code{from\_dict}. Otherwise, random problem instances cannot be saved.
\end{itemize}

While this interface leads to additional development effort for new problem classes, this object-oriented approach is necessary for the QDT to scale and handle new optimization problems automatically. Nodes can still be modified to perform tasks specific to a problem class. For instance, an instance of the Capacitated Vehicle Routing Problem (CVRP) can be decomposed into  a set of TSPs.

\subsubsection{Solvers}

The Solver class defines a common interface for algorithms. It requires an optimization problem (e.g, a QUBO matrix) and returns a result object. The form of the optimization problem and the result object can be specific to each solver. However, some decision tree nodes (e.g., a \code{SolverSetupNode}) should interpret the result on the backward pass. By convention, variational quantum solvers return the best quantum circuit they found and an instance of a Qiskit result class such as \code{SamplingMinimumEigensolverResult}. If available, information about the performance of the solver can be included in the result (processing times, the number of quantum circuit executions etc.).

A Solver object needs to implement two methods:

\begin{itemize}
    \item \code{solve}: The central function that will be called by the \code{AlgorithmExecuteNode} (see \cref{fig:decisiontree}).
    \item \code{check\_input}: Should return a boolean value so the QuaST decision tree can automatically check whether the problem data is ready to be handed over to the solver. E.g., variational algorithms may require an Ising operator to be passed as the cost function, whereas classical routines can sample from a QUBO matrix directly.
\end{itemize}

\section{Demonstration}
\label{sec:demo}

In spirit, the QDT follows a user-centered design approach. The two main user types are the end user and the developer. End users are expected to use the integrated version of the QDT via \code{python -m decisiontree} after which they can load their optimization problem and are guided through the actual decision tree. At the highest degree of automation, the QDT can be configured to implement it suggestions directly, requiring minimal user input. Developers act as an end user when exploring and benchmarking solution paths, but can contribute to the QDT by expanding it. Naturally, the behavior triggered by \code{python -m decisiontree} can be seamlessly loaded into ordinary python scripts (or Jupyter notebooks) which allows for the integration in specialized workflows including some problem generation from real-world data, or manual modification of the results.

\subsection{The End User Workflow}

The end user is an expert in their application area. They do not want to dive deeper into QC technology than necessary. Their goal is to find a better solution (more efficient, faster, higher solution quality) of a given use case. At the current stage, many users from the industry simply want to evaluate what can be achieved with quantum-assisted algorithms. The QDT significantly reduces the manual work of the end user, at least for starting with a prototypical solution. Only two manual steps are required:

\begin{enumerate}
    \item Formulate the use case in the language of mathematical optimization, e.g., as a TSP.
    \item Analyze and interpret the result of the QDT.
\end{enumerate}

As an optional third ingredient, the configuration of the QDT can be changed in order to, e.g., give more power to the end user in selecting options, or choose between different strategies for encoding the constraints. The QDT is in a prototypical stage and limited in the classes of optimization problems it can handle, with full support provided for TSP and MaxCut problems as a first step.The problem instances are simply defined in a human-readable JSON format containing the necessary data to define the problem. A TSP instance, for example, is defined by its adjacency matrix, containing the distances between the cities. Additional information is not required, but can be given. For example, city locations simplify the plotting of a solution, or labeling the paths between the cities may provide information on the vehicle that is used. The latter is a good example of information that is outside of the scope of the TSP, but may be very relevant for the interpretation of the result from a business perspective.

The \code{from\_dict} method of the problem classes specifies the JSON format for the problem instances. Together with example files and tutorial notebooks for the usage of the QDT, they are a part of the software's documentation, designed for easy usage by end users. The JSON format for problem instances in principle is designed in a way that requires minimal information about the problem instance, but additional metadata can be given to facilitate the integration with other parts of the software stack or a submission system for specific use cases that takes over, e.g., the retrieval of relevant TSP instances from real-world data.

After the QDT has finished executing, several files are generated. They help to evaluate the performance of the algorithm and analyze the solution. For end users, \code{result.json} contains the most relevant information as produced by the backward pass of the QDT. Further files are generated for reproducibility and a more thorough investigation of the algorithm's performance:

\begin{itemize}
    \item \code{problem\_instance.json}: Only if a random problem instance was generated. The main purpose of this file is to explore alternative paths on the same problem instance. Requires the problem class to have a \code{to\_dict} method.
    \item \code{run\_config.json}: The configuration data of the decision tree run. A few select modes modify the configuration. The file is mainly intended for reproducibility.
    \item \code{problem\_data.json}: The full \code{problem\_data} dictionary containing all data the decision tree generated during its forward pass. Mainly intended for debugging purposes.
\end{itemize}

\subsection{Development inside the QuaST decision tree}

Developers are needed to expand the QDT. Typically, they are experts at one or more of the blocks of the decision tree (formulation, encoding, decomposition, and so on). They want to integrate methods that they think are worth exploring and require a simple, transparent integration with the solution paths of the QDT. Integrating a method in the QDT opens the door for application-centered comparison and benchmarking, and helps new ideas gain visibility. 

For the typical development workflow expanding the decision tree, we identified three principal models described in this section.

\subsubsection{Adding a Problem Class or an Optimizer}

First, one might want to add a new option to nodes that automatically discover objects of a certain type, e.g., problem classes or optimizers. For programmers familiar with object-oriented programming, the procedure should be easy. The QDT itself does not need to be modified. A new problem class requires (1) a definition of the JSON format to define a problem instance and (2) the implementation of a subclass of the abstract \code{OptimizationProblem} (see 
\cref{sec:problemclass}). When executing the decision tree afterward, the \code{LoadProblemNode} loads the new problem class and generates random instances upon request. Alternatively, it can be loaded from an instance file in JSON format.

To include an optimizer, the developer needs to (1) provide an implementation of the optimizer that is compatible with Qiskits standards and (2) create an \code{OptimizerBuilder} instance that can create the optimizer and set its relevant hyperparameters. The \code{SelectOptimizerNode} will then discover the new optimizer and present it as an option.

\subsubsection{Adding a Node}
\label{sec:dev_node}

By design, the process for adding a node is local within the QDT. A developer adding a node will only modify the nodes in its vicinity. To repeat, nodes can perform basically any task, therefore any processing step can be added easily. For instance, let us assume the following scenario: In \cref{fig:decisiontree}, the \code{mixer} block is missing. \code{SelectLayersNode} is followed immediately by \code{SelectOptimizerNode}.The QDT can only create a standard QAOA algorithm without specifying a custom mixer. We now describe the process to implement this missing functionality.

\begin{enumerate}
    \item Implement the required nodes.
    
    For maximum modularity, the process of defining a QAOA mixer is split up into two nodes. First, the \code{LoadOrGenerateMixerNode} queries the user whether the mixer should be loaded or generated from a Qiskit template. If they answer ``load'', it asks for the file path of the mixer and loads it from a qasm3 file. If the user answers ``generate'', a second node called \code{SelectMixerNode} is executed. It finds a number of templates defined in the QDT auxiliary functions and classes, defining the X-, XY and Ring mixers. Then, the user selects a mixer template and the node constructs the mixer automatically, reading the number of qubits required from the \code{problem\_data} dictionary.

    \item Connect them to their neighbors in the QDT.

    The nodes now need to be inserted into the decision tree by modifying the \code{next\_node} function of their parent nodes. In this case, the \code{next\_node} function of the \code{SelectLayersNode} is should point to \code{LoadOrGenerateMixerNode} instead of \code{SelectOptimizerNode}. The function of the \code{LoadOrGenerateMixerNode} points to the \code{SelectMixerNode} if the user chooses to generate a mixer and to \code{SelectOptimizerNode} otherwise. Finally, the \code{SelectMixerNode} connects to \code{SelectOptimizerNode}. 
\end{enumerate}

The integration of the new functionality is now complete. However, one should ensure that the \code{problem\_data} dictionary is not modified in a way that breaks other nodes. In particular, the nodes above and below in the path need to understand it. This should be ensured by test runs with different solution paths. Good heuristics to avoid breaking changes in \code{problem\_data} are:

\begin{enumerate}
    \item Do not delete dictionary entries.
    \item Do not change the type of dictionary entries. Only change the shape (of matrices, vectors etc.) if you know what other nodes use the entry.
    \item When in doubt and when possible, create new dictionary entries and modify precisely the nodes that need to use them in the rest of the QDT.
\end{enumerate}

To end this section, we make two remarks. 

First, the above example could have been implemented with a single node. However, it is advised to design nodes such that they fulfill one clearly defined, simple task only. This helps keeping the QDT clean and allows future developers to grasp the function of the nodes more quickly. Furthermore, it simplifies the insertion (and, if necessary) removal of nodes at the appropriate location.

Second, the file structure of the QDT needs to avoid circular imports. Since any node imports its successor, this translates to the requirement that no path can reenter a block after exiting it in \cref{fig:decisiontree}. Changes that require such a path should be accompanied by a refactoring of the QDT.

\subsubsection{Adding a Closed-Source Module}

With quantum technologies and quantum software becoming more mature, monetizing efforts on quantum software will increase. The scientific groups and companies that currently work towards open-source software will move some of their products to a license-based model. QC backend access, of course, already requires access tokens for most access modes. Fortunately, the inclusion of closed-source software packages and restricted content is easily achieved by the QDT. 

Developers should follow \cref{sec:dev_node} and implement a node \code{RestrictedNode} that invokes the restricted module via its API and feeds back its result to the problem data. The necessary access token (or other form of user validation) should be put into a new \code{config} entry. A default value is given in the \code{standard\_config}. Then, the previous node should read the config to distinguish whether access is permitted. Only with a valid access token, \code{RestrictedNode} is executed. 

To be safe, it is recommended to ask the user for confirmation before invoking licensed software. Most importantly, the QDT should never use resources with a payment model where each call might produce costs or exhaust a limited budget without explicit user confirmation.
\section{Conclusion and Outlook}
\label{sec:outlook}

In this paper, we describe the guiding principles and implementation of the QuaST decision tree (QDT) as well as how to access it as an end user and how to contribute as a developer. The six principles modularity, automation without restriction, locality, configurability, light-weight execution, and transparency are certainly not the only ones that could provide the base of the decision tree. For example, one could drop the ``automation without restriction'' requirement and hide options that the developers of the decision tree seem unpractical or useless. As a representative of the other guiding principles, let us list three arguments against its removal.

First, if a developer invests their time to implement a feature, they deem it worthy of exploration. Features and methods without any merit will not be implemented in the first place. In the future, the QDT might aim for quantum advantage instead of explorative use. Then, the threshold for implemented methods will change and a review of the guiding principles is in order. For now, it is unclear whether a practical quantum advantage is possible for optimization problems, and what methods have to be used. When in doubt, the search space should not be restricted. 

Second, statements such as ``method X is useless'' are inherently uncertain and bound to a specific time and knowledge level. QC is a highly active and dynamic research where investigators still disagree about the best research direction. Therefore, the safe option is to give recommendations, but not exclude solution paths. 

Third, material on QC for the general public (press statements, media coverage) is often incomplete, out of date, or biased. The QDT helps end users to gather first-hand experiences. From a pedagogical standpoint, it is important to see what does not work. For example, industry users often come with unrealistic expectations concerning the technology readiness of algorithms like the QAOA. They learn quickly that, currently, one cannot ``simply'' use it for most use cases and get results that are remotely competitive with classical solvers.

The six guiding principles we identify contribute to a robust, yet flexible software tool that can support end users and researchers alike. The high degree of modularity is crucial because no single person or research group can hope to find optimal decisions on all branches of the decision tree. Finding useful quantum-assisted methods for optimization problems is a community effort~\cite{poggel_recommending_2023}. A framework such as the QDT is needed to avoid diverging into too many disconnected research directions. In the end, the QC community shares the goal of finding better solutions for real-world applications. Efforts like hyperparameter tuning for specific algorithms, encodings and new post-processing steps should be evaluated with their impact on relevant applications in mind.

The QDT is currently in an experimental, prototypical stage. In particular, we opted for a Python-based implementation which, to some extent, violates our own light-weight execution requirement. However, code is read much more than it is written by a common computer science platitude. At this stage, it is more important that developers have quick access to a transparent code base they can build on. Most of the scientific community in QC works in Python. Additionally, the features of the QDT still undergo frequent changes, so one should not put too much effort into specific runtime optimizations. 

On the topic of frequent changes, we close this discussion with a shortlist of features that will be implemented into the QDT in the near future.
\balance
\begin{itemize}
    \item Batch handling: The ability to load multiple problem instances of a problem class at once. A possible way is to allow a single JSON file to contain multiple problem instances.
    \item Asynchronous execution: At the end of a decision tree run, the decision tree should be able to submit jobs and execute solvers at a later time. To be implemented, this requires either saving the problem data, or using the non-interactive mode to preserve the user input. A simple, yet inefficient way is to serialize problem data to a file such as \code{problem\_data.pkl}, together with the current node. Computation could then be resumed via \code{python -m decisiontree resume problem\_data.pkl}.
    \item Interface nodes: For further modularization, the QDT will implement \emph{interface nodes} that, e.g., take over the submission to a quantum hardware provider where the compilation and transpilation can be executed. In particular, these nodes extracts the required data and packages it in a compatible way. 

In conclusion, the QDT provides a recipe and a demonstration of the upper part of a QC software stack for optimization. In particular, it takes over the extremely diverse and challenging task of systematically handling input in terms of different problem classes, formulations and algorithms with a diverse distribution of the computational load on classical and quantum systems. For end users, it is a foundational building block for a full solution pipeline with quantum-assisted algorithms. 
\end{itemize}
\clearpage
\printbibliography

@inproceedings{poggel_recommending_2023,
	title = {Recommending {Solution} {Paths} for {Solving} {Optimization} {Problems} with {Quantum} {Computing}},
	url = {http://dx.doi.org/10.1109/QSW59989.2023.00017},
	doi = {10.1109/qsw59989.2023.00017},
	booktitle = {2023 {IEEE} {International} {Conference} on {Quantum} {Software} ({QSW})},
	publisher = {IEEE},
	author = {Poggel, Benedikt and Quetschlich, Nils and Burgholzer, Lukas and Wille, Robert and Lorenz, Jeanette Miriam},
	month = jul,
	year = {2023},
}

@misc{volpe_predictive_2024,
	title = {A {Predictive} {Approach} for {Selecting} the {Best} {Quantum} {Solver} for an {Optimization} {Problem}},
	url = {https://arxiv.org/abs/2408.03613},
	author = {Volpe, Deborah and Quetschlich, Nils and Graziano, Mariagrazia and Turvani, Giovanna and Wille, Robert},
	year = {2024},
	note = {\_eprint: 2408.03613},
}

@misc{volpe_towards_2024,
	title = {Towards an {Automatic} {Framework} for {Solving} {Optimization} {Problems} with {Quantum} {Computers}},
	url = {https://arxiv.org/abs/2406.12840},
	author = {Volpe, Deborah and Quetschlich, Nils and Graziano, Mariagrazia and Turvani, Giovanna and Wille, Robert},
	year = {2024},
	note = {\_eprint: 2406.12840},
}

@misc{eichhorn_hybrid_2024,
	title = {Hybrid {Meta}-{Solving} for {Practical} {Quantum} {Computing}},
	url = {https://arxiv.org/abs/2405.09115},
	author = {Eichhorn, Domenik and Schweikart, Maximilian and Poser, Nick and Fiand, Frederik and Poggel, Benedikt and Lorenz, Jeanette Miriam},
	year = {2024},
	note = {\_eprint: 2405.09115},
}

@misc{matteo_abstraction_2024,
	title = {An {Abstraction} {Hierarchy} {Toward} {Productive} {Quantum} {Programming}},
	url = {https://arxiv.org/abs/2405.13918},
	author = {Matteo, Olivia Di and Núñez-Corrales, Santiago and Stęchły, Michał and Reinhardt, Steven P. and Mattson, Tim},
	year = {2024},
	note = {\_eprint: 2405.13918},
}

@misc{quetschlich_mqt_2023,
	title = {{MQT} {Predictor}: {Automatic} {Device} {Selection} with {Device}-{Specific} {Circuit} {Compilation} for {Quantum} {Computing}},
	url = {https://arxiv.org/abs/2310.06889},
	author = {Quetschlich, Nils and Burgholzer, Lukas and Wille, Robert},
	year = {2023},
	note = {\_eprint: 2310.06889},
}

@misc{henderson_demonstration_2024,
	title = {Demonstration of a {Hardware}-{Independent} {Toolkit} for {Automated} {Quantum} {Subcircuit} {Synthesis}},
	url = {https://arxiv.org/abs/2309.01028},
	author = {Henderson, Elena R. and Henderson, Jessie M. and Sinha, Aviraj and Larson, Eric C. and Thornton, Mitchell A.},
	year = {2024},
	note = {\_eprint: 2309.01028},
}

@misc{bandic_full-stack_2022,
	title = {Full-stack quantum computing systems in the {NISQ} era: algorithm-driven and hardware-aware compilation techniques},
	copyright = {Creative Commons Attribution 4.0 International},
	url = {https://arxiv.org/abs/2204.06369},
	publisher = {arXiv},
	author = {Bandic, Medina and Feld, Sebastian and Almudever, Carmen G.},
	year = {2022},
	doi = {10.48550/ARXIV.2204.06369},
}

@misc{bieberich_bridging_2023,
	title = {Bridging {HPC} and {Quantum} {Systems} using {Scientific} {Workflows}},
	url = {https://arxiv.org/abs/2310.03286},
	author = {Bieberich, Samuel T. and Maheshwari, Ketan C. and Wilkinson, Sean R. and Date, Prasanna and Suh, In-Saeng and Silva, Rafael Ferreira da},
	year = {2023},
	note = {\_eprint: 2310.03286},
}

@misc{saurabh_conceptual_2023,
	title = {A {Conceptual} {Architecture} for a {Quantum}-{HPC} {Middleware}},
	url = {https://arxiv.org/abs/2308.06608},
	author = {Saurabh, Nishant and Jha, Shantenu and Luckow, Andre},
	year = {2023},
	note = {\_eprint: 2308.06608},
}

@misc{chakraborty_empowering_2024,
	title = {Empowering the {Quantum} {Cloud} {User} with {QRIO}},
	url = {https://arxiv.org/abs/2407.17676},
	author = {Chakraborty, Shmeelok and Hou, Yuewen and Chen, Ang and Ravi, Gokul Subramanian},
	year = {2024},
	note = {\_eprint: 2407.17676},
}

@misc{mao_q-gen_2024,
	title = {Q-gen: {A} {Parameterized} {Quantum} {Circuit} {Generator}},
	url = {https://arxiv.org/abs/2407.18697},
	author = {Mao, Yikai and Shresthamali, Shaswot and Kondo, Masaaki},
	year = {2024},
	note = {\_eprint: 2407.18697},
}

@article{farhi_quantum_2014,
	title = {A {Quantum} {Approximate} {Optimization} {Algorithm}},
	copyright = {arXiv.org perpetual, non-exclusive license},
	url = {https://arxiv.org/abs/1411.4028},
	author = {Farhi, Edward and Goldstone, Jeffrey and Gutmann, Sam},
	year = {2014},
	doi = {10.48550/ARXIV.1411.4028},
}

@misc{kulshrestha_qarchsearch_2023,
	title = {{QArchSearch}: {A} {Scalable} {Quantum} {Architecture} {Search} {Package}},
	url = {https://arxiv.org/abs/2310.07858},
	author = {Kulshrestha, Ankit and Lykov, Danylo and Safro, Ilya and Alexeev, Yuri},
	year = {2023},
	note = {\_eprint: 2310.07858},
}

@article{peruzzo_variational_2014,
	title = {A variational eigenvalue solver on a photonic quantum processor},
	volume = {5},
	issn = {2041-1723},
	url = {https://doi.org/10.1038/ncomms5213},
	doi = {10.1038/ncomms5213},
	number = {1},
	journal = {Nature Communications},
	author = {Peruzzo, Alberto and McClean, Jarrod and Shadbolt, Peter and Yung, Man-Hong and Zhou, Xiao-Qi and Love, Peter J. and Aspuru-Guzik, Alán and O’Brien, Jeremy L.},
	month = jul,
	year = {2014},
	pages = {4213},
}

@article{palubeckis_multistart_2004,
	title = {Multistart {Tabu} {Search} {Strategies} for the {Unconstrained} {Binary} {Quadratic} {Optimization} {Problem}},
	volume = {131},
	issn = {1572-9338},
	url = {https://doi.org/10.1023/B:ANOR.0000039522.58036.68},
	doi = {10.1023/B:ANOR.0000039522.58036.68},
	number = {1},
	journal = {Annals of Operations Research},
	author = {Palubeckis, Gintaras},
	month = oct,
	year = {2004},
	pages = {259--282},
}

@article{blekos_review_2024,
	title = {A review on {Quantum} {Approximate} {Optimization} {Algorithm} and its variants},
	volume = {1068},
	issn = {0370-1573},
	url = {https://www.sciencedirect.com/science/article/pii/S0370157324001078},
	doi = {https://doi.org/10.1016/j.physrep.2024.03.002},
	journal = {Physics Reports},
	author = {Blekos, Kostas and Brand, Dean and Ceschini, Andrea and Chou, Chiao-Hui and Li, Rui-Hao and Pandya, Komal and Summer, Alessandro},
	year = {2024},
	pages = {1--66},
}

@article{tilly_variational_2022,
	title = {The {Variational} {Quantum} {Eigensolver}: {A} review of methods and best practices},
	volume = {986},
	issn = {0370-1573},
	url = {https://www.sciencedirect.com/science/article/pii/S0370157322003118},
	doi = {https://doi.org/10.1016/j.physrep.2022.08.003},
	journal = {Physics Reports},
	author = {Tilly, Jules and Chen, Hongxiang and Cao, Shuxiang and Picozzi, Dario and Setia, Kanav and Li, Ying and Grant, Edward and Wossnig, Leonard and Rungger, Ivan and Booth, George H. and Tennyson, Jonathan},
	year = {2022},
	pages = {1--128},
}

@article{cerezo_variational_2021,
	title = {Variational quantum algorithms},
	volume = {3},
	issn = {2522-5820},
	url = {https://doi.org/10.1038/s42254-021-00348-9},
	doi = {10.1038/s42254-021-00348-9},
	number = {9},
	journal = {Nature Reviews Physics},
	author = {Cerezo, M. and Arrasmith, Andrew and Babbush, Ryan and Benjamin, Simon C. and Endo, Suguru and Fujii, Keisuke and McClean, Jarrod R. and Mitarai, Kosuke and Yuan, Xiao and Cincio, Lukasz and Coles, Patrick J.},
	month = sep,
	year = {2021},
	note = {Number: 9},
	pages = {625--644},
}

@misc{abbas_quantum_2023,
	title = {Quantum {Optimization}: {Potential}, {Challenges}, and the {Path} {Forward}},
	url = {https://arxiv.org/abs/2312.02279},
	author = {Abbas, Amira and Ambainis, Andris and Augustino, Brandon and Bärtschi, Andreas and Buhrman, Harry and Coffrin, Carleton and Cortiana, Giorgio and Dunjko, Vedran and Egger, Daniel J. and Elmegreen, Bruce G. and Franco, Nicola and Fratini, Filippo and Fuller, Bryce and Gacon, Julien and Gonciulea, Constantin and Gribling, Sander and Gupta, Swati and Hadfield, Stuart and Heese, Raoul and Kircher, Gerhard and Kleinert, Thomas and Koch, Thorsten and Korpas, Georgios and Lenk, Steve and Marecek, Jakub and Markov, Vanio and Mazzola, Guglielmo and Mensa, Stefano and Mohseni, Naeimeh and Nannicini, Giacomo and O'Meara, Corey and Tapia, Elena Peña and Pokutta, Sebastian and Proissl, Manuel and Rebentrost, Patrick and Sahin, Emre and Symons, Benjamin C. B. and Tornow, Sabine and Valls, Victor and Woerner, Stefan and Wolf-Bauwens, Mira L. and Yard, Jon and Yarkoni, Sheir and Zechiel, Dirk and Zhuk, Sergiy and Zoufal, Christa},
	year = {2023},
	note = {\_eprint: 2312.02279},
}

@book{applegate_david_l_traveling_2007,
	title = {The {Traveling} {Salesman} {Problem}: {A} {Computational} {Study}},
	isbn = {978-1-4008-4110-3},
	url = {http://ieeexplore.ieee.org/document/9453454},
	publisher = {Princeton University Press},
	author = {{Applegate, David L.} and {Bixby, Robert E.} and {Chvátal, Vašek} and {Cook, William J.}},
	year = {2007},
}

@misc{javadi-abhari_quantum_2024,
	title = {Quantum computing with {Qiskit}},
	url = {https://arxiv.org/abs/2405.08810},
	author = {Javadi-Abhari, Ali and Treinish, Matthew and Krsulich, Kevin and Wood, Christopher J. and Lishman, Jake and Gacon, Julien and Martiel, Simon and Nation, Paul D. and Bishop, Lev S. and Cross, Andrew W. and Johnson, Blake R. and Gambetta, Jay M.},
	year = {2024},
	note = {\_eprint: 2405.08810},
}

@article{amaro_filtering_2022,
	title = {Filtering variational quantum algorithms for combinatorial optimization},
	volume = {7},
	url = {https://doi.org/10.1088%2F2058-9565%2Fac3e54},
	doi = {10.1088/2058-9565/ac3e54},
	number = {1},
	journal = {Quantum Science and Technology},
	author = {Amaro, David and Modica, Carlo and Rosenkranz, Matthias and Fiorentini, Mattia and Benedetti, Marcello and Lubasch, Michael},
	month = jan,
	year = {2022},
	note = {Publisher: IOP Publishing},
	pages = {015021},
}

@misc{turati_benchmarking_2023,
	title = {Benchmarking {Adaptative} {Variational} {Quantum} {Algorithms} on {QUBO} {Instances}},
	url = {https://arxiv.org/abs/2308.01789},
	author = {Turati, Gloria and Dacrema, Maurizio Ferrari and Cremonesi, Paolo},
	year = {2023},
	note = {\_eprint: 2308.01789},
}

@misc{rohe_problem_2024,
	title = {From {Problem} to {Solution}: {A} general {Pipeline} to {Solve} {Optimisation} {Problems} on {Quantum} {Hardware}},
	url = {https://arxiv.org/abs/2406.19876},
	author = {Rohe, Tobias and Grätz, Simon and Kölle, Michael and Zielinski, Sebastian and Stein, Jonas and Linnhoff-Popien, Claudia},
	year = {2024},
	note = {\_eprint: 2406.19876},
}

\end{document}